# DDoS Attacks in Cloud Computing: Detection and Prevention

Zain Ahmad
Dept. of Computer Science
FAST-NUCES
Lahore, Pakistan
l201011@lhr.nu.edu.pk

Musab Ahmad
Dept. of Computer Science
FAST-NUCES
Lahore, Pakistan
l200994@lhr.nu.edu.pk

Bilal Ahmad
Dept. of Computer Science
FAST-NUCES
Lahore, Pakistan
l201030@lhr.nu.edu.pk

*Abstract*—DDoS attacks are one of the most prevalent and harmful cybersecurity threats faced by organizations and individuals today. In recent years, the complexity and frequency of DDoS attacks have increased significantly, making it challenging to detect and mitigate them effectively. The study analyzes various types of DDoS attacks, including volumetric, protocol, and application layer attacks, and discusses the characteristics, impact, and potential targets of each type. It also examines the existing techniques used for DDoS attack detection, such as packet filtering, intrusion detection systems, and machine learning-based approaches, and their strengths and limitations. Moreover, the study explores the prevention techniques employed to mitigate DDoS attacks, such as firewalls, rate limiting , CPP and ELD mechanism. It evaluates the effectiveness of each approach and its suitability for different types of attacks and environments. In conclusion, this study provides a comprehensive overview of the different types of DDoS attacks, their detection, and prevention techniques. It aims to provide insights and guidelines for organizations and individuals to enhance their cybersecurity posture and protect against DDoS attacks.

## I. INTRODUCTION

The prevalence of Distributed Denial-of-Service (DDoS) attacks in recent years has become a growing concern for online services, jeopardizing their availability and security. In particular, cloud service providers and their customers are increasingly vulnerable to DDoS attacks due to the popularity of cloud computing. Thus, the objective of this study is to examine the impact of DDoS attacks on cloud computing. Specifically, we will investigate the characteristics of different types of DDoS attacks, including exhaustive/volumetric and protocol-based attacks. Afterwards, the study focuses towards the detection of DDoS attacks. Detection is also a key aspect of mitigating the damage caused by a DDoS attack. Traditional detection techniques are inadequate against newer attacks that target the Application Layer. Therefore, we will explore novel detection techniques leveraging AI and ML in Section IV. In Section V, we will discuss preventive measures, such as firewalls and ELD mechanisms, to stop attackers from using hosts as agents. Finally, we will conduct a comparative analysis of the proposed preventive measures in research papers to identify the most effective defence mechanisms for DoS attacks in cloud computing models. This study provides insights into the challenges and solutions related to DDoS attacks in cloud computing, which can help organizations develop robust strategies to protect their online services.

## II. LITERATURE REVIEW

DDoS attacks on cloud computing have become a serious concern in recent years, as they can cause significant disruptions to cloud-based services and pose a threat to the security of sensitive data. In this literature review, we will examine some of the key research articles on DDoS attacks on cloud computing, focusing on the different types of attacks, the impact of such attacks, and the various techniques used to mitigate them.

### A. Types of DDoS Attacks

These attacks may have thousands of versions because of the unique coding style of each attacker. Approach is usually to exploit a vulnerability of the cloud infrastructure or in case of botnets even the cloud network.[1] Botnet being a relatively modern technique for DDoS allows the attacker to be hidden because it adds a layer of nodes between the victim and the attacker which makes it difficult to trace them. Such nodes with spoofed IP addresses are called "Zombies".[2]

One of the most common types of DDoS attacks on cloud computing is the volumetric attack, which floods the network with a massive amount of traffic, making it difficult for legitimate users to access the cloud-based service. Another type of attack is the protocol attack, which exploits weaknesses in the communication protocols used by the cloud-based service.[2]

### B. Detection of DDoS Attacks

Detection is one of the most crucial steps in mitigating the effects of any DDoS attacks. Traditionally, these techniques included signature matching, threshold techniques however due to attackers targeting the application layer instead of the transport layer, these techniques have been rendered useless as packets on the application layer are difficult to identify as attack packets. When they are recognized, they can again be masked by changing their signature like IP address, port numbers etc.

Based on these new methods for detection have been proposed.[3] The use of MapReduce operations to speed up network analysis to detect any potential DDoS attacks much more efficiently and quickly. Despite the speedup offered, the algorithm still uses threshold-based detection which is prone to failure when attacker sends less packets than the threshold. The researchers have shown an accuracy of 74% when using this technique to detect DDoS packets.

Other techniques overcome these shortcomings by using AI and ML [4-5] to detect DDoS packets. This proposed method utilizes a Long-Short-Term-Memory (LTSM) Neural Network to detect DDoS packets. To further improve the accuracy of this neural network, the researchers propose the use of an optimization algorithm which combine Harris Hawks optimization and Particle Swarm optimization techniques to find the best possible combination of attributes for the LTSM. The researchers report an impressive accuracy of around 98% which is better than most other detection algorithms available.

*C. Prevention of DDoS Attacks*

DDoS (Distributed Denial of Service) attacks are a significant threat to cloud computing models, as they can cause prolonged service disruptions and can be difficult to detect and mitigate. Therefore, preventing DDoS attacks is crucial for maintaining the availability and reliability of cloud-based services. Prevention is the first line of defense against DoS attack, which ensures that the hackers cannot gain access to the system and use hosts as agents. In the research, papers referenced there are multiple defense mechanisms proposed some commonly used in the network layer such as firewalls while some are techniques, which are more modern such as ELD Mechanism. The second research paper [6] also discusses multiple techniques, which are used in the application layer as well such as rate limiting and focuses on preventive measures for specific DoS attacks. Both Papers provide valuable insight on how to prevent such attacks using modern defense mechanisms.

### III. DDoS ATTTACK TYPES

*A. Threat to Cloud Computing*

Cloud computing has become increasingly important in today's working world due to the benefits it provides to businesses of all sizes. Cloud Computing enables scalability for companies when it comes to their IT resources; it promotes flexibility in working environments; it is more cost-effective and leads to stronger collaboration between employees. All this means that in today's world cloud computing is a part of all our lives. There are different service models of cloud computing, including Platform as a Service (PaaS), Infrastructure as a Service (IaaS), and Software as a Service (SaaS). SaaS allows users to run software or applications without installing them on their own machine, while IaaS uses virtualization technology to share hardware with multiple tenants. Virtualization is extremely integral to cloud computing, as it allows for more efficient use of available hardware and can increase system availability while reducing costs. Given its advantages cloud computing is widely used and depended upon. This means any hinderance in the cloud service could have monumental effects to businesses and workflows.

DoS (Denial of Service) Attack is a major cybercrime that plagues cloud computing. DoS attacks can greatly disrupt or fully breakdown the network connectivity of a victim. Attackers compromise many hosts by launching attacks and depleting the target network's resources, with the goal of making the victim unable to use resources like web servers, CPU, storage, and other network resources. DoS attacks can significantly reduce the performance of cloud services by damaging virtual servers. Attackers can exploit vulnerabilities in software implementation or deplete all the bandwidth or resources of the victim's system.

*B. DDoS (Distributed Denial of Service)*

Unlike a regular DoS attack where a single computer sends a large amount of traffic to the target system, a DDoS attack involves multiple computers from different locations. These computers are often part of a botnet, which is a network of computers that are controlled remotely by the attacker. The attacker can then use this network to launch the attack without being easily traced. The internet's design makes it difficult to trace the attack source, as attackers use zombie machines with spoofed IP addresses. In a DDoS attack the attacker generally sends its orders to a system called the command-and-control server (C&C server) that coordinates and controls a botnet (collection of compromised hosts). The botnet obscures the attacker by providing a level of in direction.[6]

*C. Modern DDoS Attacks*

Now a days botnets are generally used to launch DDoS attacks. A botnet comprises of a Master, Handler, Agents, and a victim. The bots or agents carry out the attack on the victim's system, and the attacker communicates with them through handlers. The attacker uses various scanning techniques like Random Scan, Hitlist Scan, Route-based Scan, Divide-and-conquer Scan, Permutation Scan, Local Preference Scan, and Topological Scan to find vulnerable machines to compromise and add to the botnet. Once a host is found, the attacker searches for vulnerabilities to gain control.[6] Common Vulnerabilities and Exposures is a source of information on vulnerabilities. The larger the number of bots in the botnet, the more disruptive the attack will be. These modern attacks are more difficult to detect and prevent because the origin of a DDoS attack may even be a cloud. If the attacker fails to saturate the cloud infrastructure, then the attacker may attempt to saturate the cloud servers.[6]

In many cases these attacks generally target Bandwidth, Connection, Resources, Protocol Vulnerabilities, Data, Process flow and Infrastructure.

*D. Classification*

There are a variety of DDoS attacks sprouting in the computing world. The major types include Bandwidth based and resource-based attacks. Both types consume the entire bandwidth and resources of the network that's been exploited. Depending upon the exploited vulnerability it can be further divided into different types.

**Bandwidth Depletion Attacks:** consumes the bandwidth of the victim's network and slows down the legitimate traffic from accessing the network.[6] The bandwidth depletion attacks are commonly of two types:

Flood Attacks: The flood attack is a type of network attack where an attacker sends a large number of packets to a victim system with the help of zombies. The packets are sent using UDP or ICMP protocols, which saturate the victim's network bandwidth and slow down the system, preventing legitimate traffic from accessing the network. In UDP flood attack, the victim system generates ICMP packets with a message "destination unreachable" when no application is found on the specified ports. In ICMP flood attack, the attacker sends a large number of ICMP ping packets to the victim system, which requires a response from the victim. The bandwidth of the victim network connections is quickly depleted without servicing the legitimate users. Fragmentation, DNS flood, VoIP flood, Media data flood, and Ping flood are variations of this attack.

Amplification Attacks: Amplification attacks exploit the broadcast address feature of internetworking devices like routers to send a large number of packets to a broadcast IP address, causing systems in the broadcast address range to reply to the victim system and resulting in malicious traffic.

Smurf and Fraggle attacks are well-known examples of this type of DDoS attack, where the former uses ICMP ECHO RESPONSE packets and the latter uses UDP echo packets to target systems that support character generation. Reflectors, which are intermediary hosts or devices that respond to packets they receive, can be used for these attacks, with the return IP-address being spoofed to the victim's system.

**Resource Depletion Attacks:** The DDoS Resource depletion attack is targeted to exhaust the victim system's resources, so that the legitimate users are not serviced. The following are the types of Resource depletion attacks:

Protocol Exploit Attacks: Protocol exploit attacks aim to consume resources of the victim system by exploiting specific features of the installed protocol. TCP SYN attacks are a common example of this type, and other examples include PUSH + ACK attacks, authentication server attacks, and CGI request attacks. For instance, before two computers can initiate a secure communication channel – they must perform a TCP handshake. A TCP handshake is a means for two parties to exchange preliminary information. A SYN packet is typically the first step of the TCP handshake, indicating to the server that the client wants to start a new channel.

In a SYN flood attack, the attacker floods the server with numerous SYN packets, each containing spoofed IP addresses. The server responds to each packet (via SYN-ACKs), requesting the client to complete the handshake. However, the client(s) never responds, and the server keeps waiting. Eventually, it crashes after waiting too long for too many responses. Reflection-based DoS is a special case of Protocol Exploitation Attack, e.g., SYN flood attack.[6]

Malformed Packed Attacks: Malformed packet attacks involve sending packets with malicious information to the victim to crash it. This can be done through IP address attacks, where the packet is wrapped with the same source and destination IP address, or through IP packet options attacks, which exploit optional fields in IP packets to create a malformed packet. By setting all quality-of-service bits to one, the victim spends additional time processing the packet, and this attack is more effective when multiple zombies are used.

**Severity of DDoS Attacks:** The rise of cloud computing has led to an increase in DDoS attacks, which are a top security threat to cloud environments. A survey by VeriSign found that 74% of respondents had experienced one or more DDoS attacks in their organizations, with 31% resulting in service disruption. As the use of cloud services increases, so will the rate of DDoS attacks. Cloud systems are designed to withstand additional workload, but flooding can affect the availability of services on the same server and raise usage bills. Neighbor attacks, where a virtual machine attacks its neighbor in the same infrastructure, can also cause financial losses and harmful effects in other servers.

IV. DDoS ATTTACK DETECTION

DDoS (Distributed Denial of Service) attack detection is the process of identifying and mitigating an attack in progress. It involves analyzing network traffic, identifying patterns that are characteristic of a DDoS attack, and taking action to block or limit the malicious traffic while allowing legitimate traffic to flow.

Detection techniques typically involve monitoring traffic for unusual patterns, such as a sudden surge in traffic from many sources, or an abnormal distribution of traffic across different parts of the network. Many DDoS detection systems also use machine learning algorithms to identify patterns that are difficult for humans to detect.

In recent times, most attackers have moved from traditional DDoS attacks, which were targeted at the Transport Layer in the network model, to newer attacks which target the Application Layer. This makes the attacks more difficult to detect because most packets at the Application layer are similar and finding differences among them is extremely time consuming and difficult. Some examples of Application-level attacks are HTTP GET Flooding, Refresh attack, SQL Injection attack, CC attack and so on.

This is why traditional approaches to detecting DDoS attacks are not usable now. DDoS detection method based on a threshold for HTTP GET Request checks to see if the packets being sent exceed a specified threshold. This is no longer viable as DDoS attack can still be achieved by sending a small number of packets targeting the host' application layer. Similarly, detection method based on signature matching also fails because bandwidth attacks like DDoS do not need to exploit software vulnerabilities to be effective thus making it relatively easy for attackers to vary the type and content of attack traffic effectively changing the signature of their packets, which makes it difficult to design accurate signatures matching algorithms for DoS attacks.

This paper discusses two alternate approaches to detect DDoS attacks. These are as follows.

*A. MapReduce*

This algorithm is proposed by Choi, J., Choi, C., Ko, B., Choi, D., & Kim, P. in their paper Detecting Web based DDoS Attack using MapReduce operations in Cloud Computing Environment. Published at the Journal of Internet Services and Information Security [7].

MapReduce is a programming model and software framework for processing large amounts of data in parallel on a large cluster of commodity hardware. The MapReduce framework consists of two main components: the Map function and the Reduce function. The Map function takes input data and processes it to produce a set of intermediate key-value pairs. The Reduce function then takes these intermediate key-value pairs and combines them into a smaller set of output key-value pairs. MapReduce can be used to detect DDoS attacks in a cloud computing environment by processing log files faster. During a DDoS attack, the datasets generated are voluminous and analyzing them for a possible attack can take hours.

MapReduce can facilitate faster processing of these log files by dividing a log file into multiple smaller chunks and processing each chunk separately over a cluster node in parallel [8].

The steps followed to detect DDoS attack by using the MapReduce framework are as follows:

- **Define a threshold for GRPS**: Normal systems do not send GET requests to the same page over and over again thus this parameter can help identify any anomalies in the network. Care is needed when setting a threshold as a low threshold can lead to false positives and high threshold can lead to false negatives.

- **Collect and preprocess network traffic data**: Uses packet loader and packet collector modules. The packet loader stores collected packets in HDFS, while the packet collector gathers packet information using Libcap and Jpcap modules from a live interface.

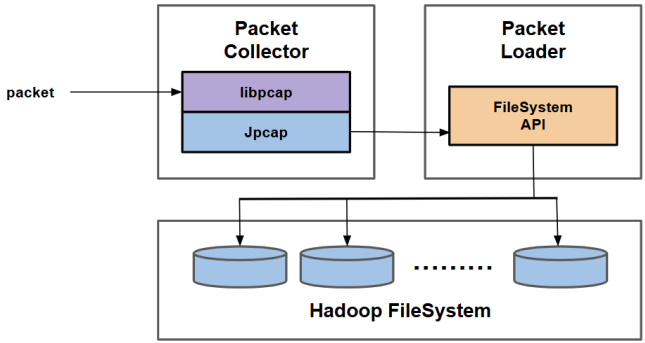

Fig. 2. Packet Capturing for Map-Reduce Algorithm.

- **Implement MapReduce jobs**: MapReduce jobs can be implemented to analyze the preprocessed network traffic data. The jobs can be designed to extract relevant features from the data, such as packet size, source IP address, destination IP address, and protocol type. This is used to extract various parameters like CPU usage, Load, packet size, information distribution of packet header, protocol distribution for classification distribution of network service, the maximum value, minimum value of traffic, monitoring of flow using spoofing address and so on which will be used in network analysis.

- **Analyze the data using statistical methods**: The extracted features can be analyzed using statistical methods and then compared with the threshold value set by the user earlier.

**Results**: The algorithm was tested on a computer system consisting of 8 nodes with one master node. The master node performs the MapReduce operation by dividing the workload on other 8 nodes. Each system is connected to Gigabit ethernet. NetBot attack tool was used to create the DDoS attack. On running the experiment, the accuracy of the algorithm was found to be around 74.7% given an HTTP Get request. The percentage of false positives was 0%.

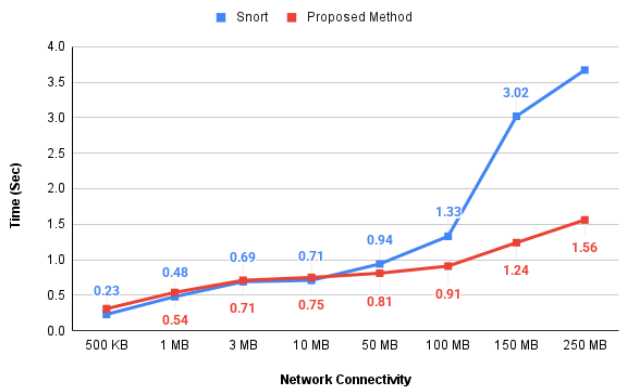

Fig. 4. Map-Reduce Algorithm Results.

## B. Neural Network and Deep Learning Based Approach

This approach is proposed by S. Sumathi, R. Rajesh, Sangsoon Lim in their paper "Recurrent and Deep Learning Neural Network Models for DDoS Attack Detection", published at the Journal of Sensors [9].

This method proposes the use of a Long Short-Term memory (LSTM) recurrent neural network and an autoencoder- and decoder-based deep learning strategy with a gradient descent learning rule to detect Distributed Denial of Service (DDoS) attacks in a cloud computing environment.

The algorithm uses a dataset consisting of 41 features to train the neural network for DDoS detection. For better optimization, the algorithm also uses Harris Hawks optimization (HHO) algorithm to select the most optimal combination of features to get the best possible accuracy. The HHO is a GA based optimization algorithm, but it suffers from slow convergence. This problem is overcome by using a hybrid algorithm consisting of Harris Hawks optimization (HHO) and Particle Swarm Optimization (PSO) for optimization.

The dataset is first fed into the HHO-PSO hybrid optimization algorithm to find the best possible combination of the features which can maximize the prediction accuracy of the LTSM. The population of features selected in each iteration of the HHO-PSO algorithm is given as

Once the algorithm has finished optimization the selected attributes are fed to the LTSM for training the neural network.

The LTSM is then trained using these attributes from the dataset which was initially divided into test and training datasets. The LTSM uses a total of 4 layers, 1 input layer, 2 hidden layers and 1 output layer. The sigmoidal activation function is used at each neuron. The learning rate is set to 0.25 and gradient descent rule is used for updating the weights in each iteration.

**Results:** The model runs 10 iterations on the NSL benchmark dataset. The results showed that the LTSM using the optimization parameters produced by the HHO-PSO algorithm produced an average accuracy of 98.5% beating out most other DDoS detection algorithms like Naïve Bayes, LSTM, MLP, and even the MapReduce algorithm discussed above.

## V. DDoS ATTACK PREVENTION

DoS (Denial of Service) attacks pose a significant security challenge for cybersecurity researchers working on cloud computing models. To mitigate such attacks, a combination of network and application layer measures can be implemented. There are three defense types available against a DoS attack: attack prevention, detection and response. In the upcoming section, we will discuss the second type of defense, which is attack prevention. Attack prevention is concerned with preventing a DoS attack to compromise, use hosts as agent, and to mitigate the effects of a DoS attack, one of the common ways to prevent Dos attack is using a firewall. Preventing DoS attacks in cloud computing models is essential for maintaining the integrity, completeness and availability of cloud-based services. We will discuss some of the defensive measures in research papers referenced and conduct a comparative analysis to determine which defenses are commonly used and which ones have been proven more effective than others.

*A. DDOS Attack Prevention Techniques*

**Firewall:** The most used defense mechanism to prevent DDoS attacks is the use of firewalls. Firewalls were once thought to be the first layer of defense against DDoS attacks. Firewalls have a set of rules set by the network administrator, which are used to inspect each incoming, and outgoing packets. Network traffic is compared with a set of predefined rules and patterns to identify any anomalies, which can indicate an attack. Only packets, which fall within a certain criterion, can pass through. Packets seen as malicious are discarded and the IP address of a user deemed to be sending these malicious packets is blocked therefore preventing the intruder from flooding the network with unnecessary packets.

However, the problem with using firewalls are that a hacker can send superfluous packets in order to degrade its functionality or they can create IP addresses and ports for packets to manipulate the firewall. Attacks like these allow the intruders to compromise the target's network bandwidth and services, making it easier to prevent access to authorized users [10].

**ELD Mechanism:** ELD is a defense mechanism that specifically targets DoS attacks, which exploit vulnerabilities in communication protocols. By analyzing the entropy or randomness of network traffic, it can identify anomalous patterns that indicate the presence of a DoS attack. This mechanism assumes that such attacks generate a large number of packets that are identical or similar in terms of their protocol headers. Compared to other defense mechanisms like firewalls, ELD is a lightweight technique that does not require a significant amount of processing power or resources.

Results have shown that implementing ELD leads to significant improvements in the true positive rate while also reducing the number of false alarms, resulting in a more efficient and effective detection of DoS attacks. Moreover, ELD can substantially decrease the cost associated with the controller [10]. However, ongoing research is being conducted to further improve the accuracy and effectiveness of this technique [10].

In summary, ELD is a protocol-based defense mechanism that can be used to detect and prevent DoS attacks in cloud computing models. By analyzing the randomness of network traffic, it can identify and prevent anomalous patterns that may indicate the presence of a DoS attack, making it an effective and efficient solution for DoS attack prevention.

**Rate Limiting:** Rate limiting is a widely used technique for preventing Denial of Service (DoS) attacks by controlling the rate of incoming traffic from an IP address. The goal is to limit the number of requests or connections an IP address can make to a server or network resource within a specific time period, thus preventing the network from becoming overloaded with an excessive number of requests. By limiting the number of requests, rate limiting helps prevent malicious attackers from overwhelming the network by sending massive amount of packets, and prevents the attacker from flooding the firewall with superfluous packets, which could potentially cause a network or server outage.

Rate limiting can be implemented at different layers of the network, including the application, network, and transport layer, depending on the specific needs of the system. At the application layer, rate limiting can be used to limit the number of requests a user can make to a web application or API within a given timeframe. At the network layer, rate limiting can be applied to restrict the number of packets an IP address can send to a particular server or resource. Similarly, at the transport layer, rate limiting can be used to control the rate of incoming traffic to a particular port.

However, it is worth noting that rate limiting alone may not always be sufficient to prevent DoS attacks, as attackers can use various techniques to bypass rate limiting measures. For instance, they may use multiple IP addresses or employ other sophisticated techniques to evade rate limiting. Therefore, it is essential to combine rate limiting with other preventive measures, such as IP blocking, traffic filtering, and behavior analysis, to enhance the security posture of the system and prevent DoS attacks.

**Client Puzzle Protocol:** The Client Puzzle Protocol (CPP) is a defense mechanism against Denial of Service (DoS) attacks. It was proposed as a solution to address the problem of DoS attacks that exploit the computational resources of the server, thereby preventing illegitimate clients from accessing the server. The main idea behind CPP is to require the client to solve a cryptographic puzzle before the server responds to the client's request. This is done in order to ensure that the client is a legitimate client and not a malicious attacker attempting to flood the server with requests.

The cryptographic puzzle that the client is required to solve is computationally expensive and requires a significant amount of processing power. This makes it difficult for attackers to generate a large number of requests and overwhelm the server. The server can set the difficulty of the puzzle based on the available resources and the desired level of security. If the client is able to solve the puzzle, the server responds to the client's request. If not, the server can either drop the request or ask the client to solve another puzzle.

One advantage of CPP is that it does not require any additional hardware or software. It can be implemented as a simple extension to existing protocols, such as the Transmission Control Protocol (TCP). Additionally, it does not require any modification to the client's software or configuration. The client simply needs to be able to solve the puzzle.

However, there are some potential drawbacks to CPP. One issue is that it can increase the latency or delay in responding to legitimate clients. This can be mitigated by setting the difficulty of the puzzle appropriately. Another issue is that it can be vulnerable to certain types of attacks, such as precomputation attacks, where an attacker precomputes the solutions to the puzzles in advance. Overall, CPP is a useful defense mechanism against DoS attacks and can be an effective tool in a multi-layered defense strategy.

### B. Comparative Analysis

The paper [10] is more focused on application layer attacks and provides a detailed analysis of several defense mechanisms for each attack, including their advantages and disadvantages. This thorough analysis allows for a better understanding of which defense mechanism would be more effective against a specific application layer attack.

In addition, the first research paper [10] also discusses some of the modern techniques used to prevent DoS attacks, such as Entropy-Based Lightweight DDoS Detection (ELD), Support Vector Machine (SVM) algorithm, and Fast Aggregative Multidimensional Scaling (FADM). These techniques are designed to enhance the accuracy and effectiveness of DoS attack prevention mechanisms.

Overall, both research papers offer valuable insights into preventing DoS attacks. However, the first paper [10] provides a more detailed and comprehensive analysis of application layer attacks and modern techniques for preventing DoS attacks, making it a useful resource for researchers and practitioners in the field.

## CONCLUSION

In conclusion, DDoS attacks continue to pose a significant threat to networks and online services, and detecting and preventing them is becoming more challenging as attackers develop new and more sophisticated techniques. Traditional approaches to detecting DDoS attacks based on threshold limits and signature matching are no longer viable as attacks have moved from targeting the Transport Layer to targeting the Application Layer.

Two alternate approaches to detect DDoS attacks discussed in this study are MapReduce and Neural Network and Deep Learning Based Approach. The MapReduce algorithm is a programming model and software framework for processing large amounts of data in parallel, which can facilitate faster processing of voluminous log files generated during DDoS attacks. The Neural Network and Deep Learning Based Approach use deep learning techniques such as Long Short-Term Memory (LSTM) recurrent neural networks and autoencoder-decoder-based strategies to detect DDoS attacks with high accuracy.

Preventing DDoS attacks requires a combination of techniques, including implementing firewalls, ELD mechanism, Control Puzzle Protocol and Rate Limiting. The use of these techniques allows us to mitigate and prevent DDoS attacks.

In summary, detecting and preventing DDoS attacks requires a comprehensive and multi-faceted approach that combines both traditional and advanced techniques, as well as ongoing vigilance and readiness to respond to evolving threats.

## REFERENCES

[1] K. Kaur and J. Ayoade, "Analysis of DDoS Attacks on IoT Architecture," 2023 10th International Conference on Electrical Engineering, Computer Science and Informatics (EECSI), Palembang, Indonesia, 2023, pp. 332-337, doi: 10.1109/EECSI59885.2023.10295766.

[2] panelRashmi V. Deshmukh a et al. Understanding DDoS Attack & its Effect in Cloud Environment

[3] A. N. H. D. Sai, B. H. Tilak, N. S. Sanjith, P. Suhas and R. Sanjeetha, "Detection and Mitigation of Low and Slow DDoS attack in an SDN environment," 2022 International Conference on Distributed Computing, VLSI, Electrical Circuits and Robotics ( DISCOVER), Shivamogga, India, 2022, pp. 106-111, doi: 10.1109/DISCOVER55800.2022.9974724.

[4] F. Reza, "DDoS-Net: Classifying DDoS Attacks in Wireless Sensor Networks with Hybrid Deep Learning," 2024 6th International Conference on Electrical Engineering and Information & Communication Technology (ICEEICT), Dhaka, Bangladesh, 2024, pp. 487-492, doi: 10.1109/ICEEICT62016.2024.10534545.

[5] A. Sebbar and K. Zkik, "Enhancing Resilience against DDoS Attacks in SDN -based Supply Chain Networks Using Machine Learning," 2023 9th International Conference on Control, Decision and Information Technologies (CoDIT), Rome, Italy, 2023, pp. 230-234, doi: 10.1109/CoDIT58514.2023.10284387.

[6] M. M. a. M. Jalali, "A survey and taxonomy of DoS attacks in cloud computing," Wiley Online Library, p. 28, 2016.panelRashmi V. Deshmukh a et al. Understanding DDoS Attack & its Effect in Cloud Environment

[7] Choi, J., Choi, C., Ko, B., Choi, D., & Kim, P. (2013). Detecting Web based DDoS Attack using MapReduce operations in Cloud Computing Environment. Journal of Internet Services and Information Security (JISIS), 3(3/4), 28-37.

[8] V. Maheshwari, A. Bhatia and K. Kumar, "Faster detection and prediction of DDoS attacks using MapReduce and time series analysis," 2018 International Conference on Information Networking (ICOIN), Chiang Mai, Thailand, 2018, pp. 556-561, doi: 10.1109/ICOIN.2018.8343180.

[9] S. Sumathi, R. Rajesh, Sangsoon Lim, "Recurrent and Deep Learning Neural Network Models for DDoS Attack Detection", Journal of Sensors, vol. 2022, Article ID 8530312, 21 pages, 2022. https://doi.org/10.1155/2022/8530312.

[10] M. T. ,. H. M. K. A. A. Ammarah Cheema, "Prevention Techniques against Distributed Denial of Service Attacks in Heterogenous Networks," WILEY , p. 15,